\documentclass[reqno]{article}

\usepackage{amsmath,amssymb}
\usepackage{latexsym}
\usepackage{cite}
\usepackage{graphicx}
\usepackage{epsfig}
\usepackage{fancyheadings}

\newcommand{\pcdv}{\itindex{\phi}{CDV}}

\newcommand{\uone}{\mbox{\slshape U}(1)}

\newcommand{\diff}[1][]{\mbox{d}#1}

\newcommand{\half}[1]{\ensuremath{\frac{#1}{2}}}
\newcommand{\intd}[1]{\int \!\! #1 \;}
\newcommand{\inv}[1]{\ensuremath{\frac{1}{#1}}}

\newcommand{\itindex}[2]{\ensuremath{#1_{\mbox{\scriptsize{\itshape #2}}}}}

\DeclareMathOperator{\extdm}{d}
\newcommand{\extd}{\extdm \!}
\title{Constant Dilaton Vacua and Kinks in 2D (Super-)Gravity}

\bigskip
\author{Luzi Bergamin \\
\small\it Institute for Theoretical Physics,\\[-1.mm]
\small\it Vienna University of Technology, 1040 Vienna, Austria,  \\[-1.mm]
\small\it email: bergamin@tph.tuwien.ac.at}
\date{ }

\begin{document}

\maketitle
\thispagestyle{fancy}

\rhead{TUW-05-13\\hep-th/0509183}

\bigskip

 \begin{abstract}
2D dilaton (super-)gravity contains a special class of solutions with
constant dilaton, a kink-like solution connecting two of them was
recently found in a specific model that corresponds to the KK reduced
3D Chern-Simons term. Here we develop the systematics of such
solutions in generalized 2D dilaton gravity and
 supergravity. The existence and characteristics thereof essentially
 reduce to the discussion of the conformally invariant potential $W$,
 restrictions in supergravity come from the relation $W=- w^2$. It is
 shown that all stable kink solutions allow a supersymmetric extension
 and are BPS therein. Some
 examples of polynomial potentials are presented.

\bigskip


 \end{abstract}

\section{Introduction}
\label{sec:intro}
Recently the study of the Kaluza-Klein reduced 3D gravitational
Chern-Simons term attracted attention due to an interesting kink
solution originally found in ref.\ \cite{Guralnik:2003we} (cf.\ e.g.\ refs.\
\cite{Grumiller:2003ad,Jackiw:2003fb,Bergamin:2004me}, an unexpected
relation to 4D gauged $N=2$ supergravity has been found in ref.\ \cite{Cacciatori:2004rt}). It was shown in
ref.\ \cite{Guralnik:2003we} that within a certain conformal frame the KK
reduced Chern-Simons term leads to the action
\begin{equation}
  \label{eq:intro1}
  \mathcal{S}= - \inv{8 \pi^2} \intd{\diff{^2 x}} \sqrt{-g} (F R +
  F^3)\ ,
\end{equation}
where $R$ is the curvature scalar and $F$ the dual field strength
tensor of the $\uone$ gauge field. In that paper three types of
solutions to the action \eqref{eq:intro1} were presented, the first one with $R=c$,
$F=0$, the second one with $R=-2c$, $F=\pm \sqrt{c}$ and a third one
given the interpretation of a kink connecting two solutions of the
second type with different signs of $F$. In ref.\ \cite{Grumiller:2003ad} it
was pointed out that eq.\ \eqref{eq:intro1} can be reformulated as a first
order action of generalized dilaton gravity (cf.\ \cite{Grumiller:2002nm} and refs.\
therein), which immediately allows to discuss \emph{all} classical
solutions at a global level. This also allowed a straightforward
supersymmetrization \cite{Bergamin:2004me} and it was shown that the
kink solution is a BPS state therein.

The reformulation in ref.\ \cite{Grumiller:2003ad} suggests that kink-like
solutions as found in ref.\ \cite{Guralnik:2003we} exist in many other
models of generalized 2D dilaton gravity as well. It is the purpose of
this paper to examine the questions of existence and behavior
thereof. In section \ref{sec:cdvs} we define the kink as a certain
solution of generalized dilaton gravity and discuss its basic
properties in the bosonic model as well as within the supersymmetric
extension. The stability of these solutions is
addressed in section \ref{sec:stability} and  in
section \ref{sec:polynom} polynomial
potentials are discussed more in detail. Our conclusions are
presented in section
\ref{sec:conclusions}.

\section{Constant Dilaton Vacua and Kinks}
\label{sec:cdvs}
We consider generalized 2D dilaton gravity in its first order
formulation with the action\footnote{The dilaton $\phi$ and $X^a$ are
  scalar fields, $\omega$ and $e_a$ are the \emph{independent} spin
  connection and the zweibein, resp., the latter yields the volume form $\epsilon = \epsilon^{ab}e_b
  \wedge e_a$. We work in light-cone coordinates labeled as ``$++$''
  and ``$--$''. The notations and conventions used here are equivalent
  to the ones used in refs.\ \cite{Ertl:2000si,Bergamin:2003am}.}  (cf.\ ref.\ \cite{Grumiller:2002nm})
\begin{equation}
  \label{eq:1.1}
  \mathcal{S} = \int_M \phi \extd \omega + X^a D e_a + \epsilon \bigl(
  V(\phi) + X^{++} X^{--} Z(\phi)\bigr)\ .
\end{equation}
As one of the unique advantages of this formulation its equations of motion 
\begin{align}
  \label{eq:1.2}
\extd \phi + X^{++} e_{++} - X^{--} e_{--} &= 0\ , &
(\extd \mp \omega) X^{\pm\pm} \pm (V+X^{++}X^{--}Z) e_{\mp\mp} &= 0\ , \\
\label{eq:1.4}
\extd \omega + \epsilon (V' + X^{++}X^{--} Z') &= 0\ , &
(\extd \mp \omega) e_{\pm \pm} + \epsilon X^{\mp \mp} Z &= 0
\end{align}
immediately unravel the integrability of the model.
Generic solutions \cite{Grumiller:2002nm} are obtained on a patch with $X^{++} \neq 0$ (or equiv.\
$X^{--} \neq 0$), they are parametrized
by the dilaton, $X^{++}$ ($X^{--}$) and the Casimir function (constant of motion)
\begin{align}
  \label{eq:1.6}
C &= e^Q X^{++}X^{--} + W\ , & Q&=\int_{\phi_0}^\phi \extd \varphi Z(\varphi)\ , & W&=
\int_{\phi_0}^\phi \extd \varphi e^Q V(\varphi)\ .
\end{align}
In Eddington-Finkelstein gauge the line element is obtained as
\begin{align}
  \label{eq:1.8}
  (\extd s)^2 &= 2 \extd u \extd \phi + K(\phi,C) (\extd u)^2\ , &
  K=2 e^Q(C-W) & = 2 e^Q X^{++} X^{--}\ ,
\end{align}
where $K$ is the Killing norm associated to the Killing vector
$\partial/\partial u$. This last equation also makes the double
nature of $\phi$ explicit: on the one hand it is a field, the dilaton, on the
other hand it may be interpreted as one of the world-sheet coordinates.

There exist solutions of eqs.\ \eqref{eq:1.2}-\eqref{eq:1.4} with
$X^{++} = X^{--} = 0$. If they do not occur at isolated points (bifurcation points)
the dilaton must be constant and thus they are called ``constant dilaton
vacua'' (CDV).  As can be seen from eq.\
\eqref{eq:1.2}, such CDVs do not exist generically but at
the roots of the potential $V$ only, $V(\pcdv)=0$. The first equation
in \eqref{eq:1.4} tells us that the curvature is constant as well, as
\begin{align}
  \label{eq:1.7}
  \extd \omega &= - \epsilon V' (\pcdv)\ , & R &= - 2 \ast \extd \omega = 2
  V'(\pcdv)\ .
\end{align}
The geometric structure of CDVs is thus AdS space ($R<0$), dS space
($R>0$) or Minkowski or Rindler space ($R=0$). Finally the Casimir
function is simply the value of the conformally invariant potential
$W(\pcdv)$. Notice that the value of $Z$ (the conformal frame) is
irrelevant and thus CDVs do not change under globally regular
conformal transformations.

\subsection{Kinks in Bosonic Models}
\label{sec:nonsusy}
The discussion of CDVs is important in the current context, as the
three solutions with constant curvature of the action \eqref{eq:intro1} found in
ref.\ \cite{Guralnik:2003we} exactly
correspond to the three CDVs of the potential \cite{Grumiller:2003ad}
\begin{align}
\label{eq:1.7.1}
V &\propto \phi^3 - c \phi\ \ \ (c>0)\ , & \phi_1 &= -\sqrt{c}\ , & \phi_2 &=
0\ , & \phi_3 &= \sqrt c\ .
\end{align}
The kink solution belongs to the class of generic solutions discussed above and
approaches the CDV solutions at the points $\phi=\phi_1$ and $\phi=\phi_3$.

We generalize this concept to arbitrary potentials $V(\phi)$ with CDVs $\phi_1<\phi_2\ldots<\phi_n$
and define the kink as a solution of \eqref{eq:1.2}-\eqref{eq:1.4} that
approaches the CDV solutions at the points $\phi_1$ (the initial point
$\phi_I$) and $\phi_n$ (the final point $\phi_F$),
i.e. $\itindex{X^{++}}{kink}$ and $\itindex{X^{--}}{kink}$ vanish at
these points. The remaining CDV points $\phi_2\ldots\phi_{n-1}$ lie in the
``interior'' of the kink and the $\itindex{X^a}{kink}$ do not
necessarily vanish there. 

To discuss the systematics of such kinks we have to address the
question how different CDVs occur and how a kink solution connecting two of them looks like. The
key observation is that this basically boils down to a discussion of
the function $W(\phi)$. As long as the conformal factor $\exp( Q)$ remains regular and non-zero
(what we will assume in the following) CDVs occur at local extrema or
turning points of $W$ with
\begin{align*}
  W''>0 & \Leftrightarrow \text{dS space,} &
  W''<0 &\Leftrightarrow\text{AdS space,}&
  W''=0 &\Leftrightarrow\text{Minkowski or Rindler space.}
\end{align*}
The value of the Casimir function is of interest
as well. As it is a constant of motion, kink
solutions only can exist between two CDVs with the same value of thereof:
$W(\phi_I) = W(\phi_F) \equiv C$. Moreover, also the Killing norm is determined by the conformally
invariant potential according to eq.\
\eqref{eq:1.8}, in a kink solution horizons occur wherever
$W(\phi_h) = W(\phi_I)$. Thus it is seen that---up to the exact value of
the curvature scalar---all
information about CDVs and possible kinks is stored in the
function $W(\phi)$. 

From these observations we can read off the following immediate consequences:
\begin{itemize}
\item An AdS and a dS CDV are never neighboring.
\item The existence of kink solutions requires at least three
  CDVs, an AdS-dS kink at least four CDVs. The simplest kink
  is AdS-dS-AdS (or dS-AdS-dS) and realized e.g.\ in the model of
  ref.\ \cite{Guralnik:2003we}.
\item At the initial and final points a kink always approaches an
  extremal horizon.
\item An AdS-AdS (dS-dS) kink always has an even number,
  an AdS-dS kink always an odd number of horizons. Here extremal horizons must be
  counted according to their multiplicity, i.e. extrema of $W$ count
  with an even number, turning points with an odd number. In particular it follows that an AdS-dS kink always has
  at least one horizon in its interior.
\item The number of horizons in the interior of a kink is limited by
  the number of CDVs, in particular
  \begin{equation}
    \label{eq:1.9}
     \#(\mbox{CDVs}) \geq \#(\mbox{NEH}) + 2  \#(\mbox{EH}) + 1\ ,
  \end{equation}
  where NEH stands for ``non-extremal horizons'' and EH for ``extremal
  horizons'', resp., and the CDVs/horizons at $\phi_I$ and $\phi_F$
  are not counted. Extremal horizons count twice as they can can occur
  at the points $\phi_2\ldots\phi_{n-1}$ only, i.e.\ an extremal
  horizon always requires the existence of a CDV with $W(\pcdv)=C$.
\end{itemize}
\subsection{Kinks in Supergravity}
\label{sec:sugra}
Supersymmetric extensions of generalized dilaton gravity have been
known for some time \cite{Park:1993sd}, its first order formulation,
i.e.\ the supersymmetric extension of the action
\eqref{eq:1.1}, has been formulated in 
refs.\ \cite{Ertl:2000si,Bergamin:2002ju,Bergamin:2003am}. As in the bosonic
model all classical solutions are easily derived therefrom.

Supersymmetry imposes a condition on the form of the potential $V(\phi)$ as
it must be expressible in terms of a prepotential $u(\phi)$, which
appears in the spinorial part of the action\footnote{The complete
  supergravity action is not presented here, as we consider solutions
  with vanishing fermionic fields, only. The complete action and its
  solutions have been presented in \cite{Bergamin:2003am}.}:
\begin{align}
  \label{eq:2.1}
  V&= - \inv{8} \bigl((u^2)' + u^2 Z  \bigr) & W &= - 2 w^2 & w&= \inv{4} e^{Q/2} u 
\end{align}
There exist two different types of CDVs,
namely\footnote{The Casimir function is defined up to an
  additive constant, only. In contrast to the bosonic model there
  exists a preferred choice in supergravity by demanding that BPS
  states have $C=0$ \cite{Bergamin:2003mh}.}
\begin{align}
  \label{eq:2.2}
  u(\pcdv)&=0=w(\pcdv) &\rightarrow&& C&=0 & R&=  -8 e^{-Q}(w')^2 \leq 0\ , \\
   (\frac{u'}{u} + \half{Z})\bigr|_{\phi=\pcdv} &= 0 = w'(\pcdv) &\rightarrow&& C&=- 2 w^2 \leq 0 & R&= - 8
  e^{-Q} w w'' \ .
\end{align}
Except for the special case $w=w'=0$, a kink always connects two $w=0$ or two $w'=0$ CDVs, mixed kinks
do not exist. Further between two $w=0$ CDVs there must be at least one
with $w'=0$.

As $W$ is bounded from above $\phi_I$ and $\phi_F$
can never be dS CDVs, which are local minima of $W$. Therefore in
supergravity dS-dS and AdS-dS kinks do not exist. This is obvious for
$w=0$ kinks but according to our definition in section \ref{sec:nonsusy}
it holds for $w'=0$ kinks as well. Further characteristics of the two
types of kinks are:
\begin{description}
\item[\boldmath $w=0$ kinks:] All (interior) horizons are maxima of $W$ and thus extremal. All
  $w=0$ CDVs are at the same time horizons and vice versa. $w=0$ kinks
  are BPS states (cf.\ refs.\ \cite{Bergamin:2003mh,Bergamin:2004me}).

\item[\boldmath $w'=0$ kinks ($w\neq0$):] They can have extremal and non-extremal horizons in its interior. The extremal
  ones are $w'=0$ CDVs. These kinks are never BPS.
\end{description}

It is interesting to consider the relation \eqref{eq:1.9} in
supergravity. For both types of kinks one obtains ($\phi_I$ and $\phi_F$ are not counted)
\begin{equation}
  \label{eq:2.3}
  \#(w=0\ \mbox{CDVs}) \leq \#(w'=0\ \mbox{CDVs}) -1\ .
\end{equation}
For BPS kinks this is a simple rephrasing of the relation
\eqref{eq:1.9} for the special case where all
horizons are extremal.

More interesting is the non-BPS kink. As
$\phi_I$ and $\phi_F$ cannot be dS, $\phi_2$ and $\phi_{n-1}$ are not
$w=0$ CDVs which yields the ``$-1$'' in \eqref{eq:2.3}. If $\phi_I$ or
$\phi_F$ are $R=0$ CDVs it is simpler to replace \eqref{eq:2.3} by
\begin{equation}
\#(w=0\ \mbox{CDVs}) \leq \#(w'=0\ \mbox{CDVs}) -3\ ,
\end{equation}
where \emph{all} CDVs are
counted according to their multiplicity. On top of this \eqref{eq:1.9}
is still a non-trivial inequality for this type of kinks.

The first order formulation allows the discussion of extended
supergravity as well, in particular the $N=(2,2)$ version as shown in
refs.\ \cite{Bergamin:2004sr,Bergamin:2004na}. This appears to yield an
especially interesting extension, as the field content contains an additional
scalar field, the scalar partner of the $\uone$ gauge field. More
specifically, one can define a complex ``dilaton'' $X=\phi + i \pi$ and
supersymmetry restricts $w(X)$ to be an analytic function in $X$
\cite{Bergamin:2004sr}, which could result in unique advantages in the
discussion of polynomial potentials. However, it turns out that all
$N=(2,2)$ kinks can be realized in the bosonic model already, as the
existence of CDV solutions does not yet imply the existence of a
kink. Indeed, the complex ``dilaton'' remains a cross: its real part
can be interpreted as a coordinate while the imaginary part is simply
a constant of motion\footnote{This is correct for bosonic field
  configurations, only. Non-trivial fermionic fields yield soul
  contributions to the charge \cite{Bergamin:2004sr}, which leads to a
subtle modification of the solutions \cite{Bergamin:2004na}.} (Casimir function), namely the charge.
Thus a kink can only exist between
two CDVs with the same charge, which in addition always can
be redefined in such a way that it vanishes on the kink
solution\footnote{In contrast to the mass $C$ there exists no
  preferred choice to fix the arbitrary constant in the definition of
  the charge, as BPS
states exist for any value thereof \cite{Bergamin:2004na}.}.

\section{Stability of the Kink}
\label{sec:stability}

If the kink shall describe the non-trivial configuration of a certain
vacuum state, we should worry besides its mere existence about its
stability as well. Two different criteria are discussed here:
thermodynamical stability and energy considerations. We start with the
thermodynamical stability, which will turn out to yield a less
restrictive bound.

All our kink solutions exhibit at least two (Killing) horizons and
thus we may ask about the Hawking effect. The Hawking temperature can
be calculated as a purely geometric quantity from surface gravity
(cf.\ e.g.\ \cite{waldgeneral}) as
\begin{equation}
  \label{eq:2.4}
  T_H = \inv{4 \pi} \bigl| \frac{\diff}{\diff \phi} K(\phi,C) \bigr|_{\phi = \phi_h} =
  \inv{2\pi} \bigl| V(\phi) \bigr|_{\phi = \phi_h}\ .
\end{equation}
Here, $\phi_h$ is the value of the dilaton at the horizon, the
prefactor is subject to conventions, but remains unimportant in the
following discussion. Not surprisingly the Hawking temperature vanishes
if $\phi_h$ is at the same time a CDV of $V$, i.e.\ if the horizon is
extremal. Therefore, the outer horizons at $\phi_I$ and $\phi_F$
always have vanishing Hawking temperature. However, this need not be true
for eventual interior horizons. But in that case we encounter a
thermodynamical instability as there are always at least two neighboring
horizons with different temperatures. Of course, these considerations
are rather academic in the current context. Our model does not have
any physical degrees of freedom and thus there cannot be any thermal
radiation. A more elaborate study of thermodynamical stability should
thus start from the kink coupled to matter fields, in that context
also a more reliable determination of the Hawking temperature is
possible (cf.\ ref.\ \cite{Grumiller:2002nm}).

As noted above, kink-like solutions are usually of interest if they
represent a non-trivial vacuum configuration. Thus we should look for
kinks that are solutions of lowest energy. A simple
definition of energy is possible in 2D dilaton gravity thanks to the
Casimir function defined in eq.\ \eqref{eq:1.6}: It is essentially equivalent to
(minus) the ADM mass wherever the latter is defined and it yields a conserved
quantity in any space-time and---with an appropriate
completion---even after the coupling of dynamical degrees of freedom
\cite{Kummer:1995qv,Grumiller:1999rz} (for supersymmetric theories
cf.\ ref.\ \cite{Bergamin:2003mh}). In our case we thus may define the
energy as
\begin{equation}
  \label{eq:2.5}
  M = -C = -W|_{\phi = \phi_h}\ .
\end{equation}
Therefore the kink is stable if $W$ is bounded from above and if
$W(\phi_h)$ represents the maximum of $W$. Then no horizon can occur
for $M<M_{\mbox{\tiny kink}}$ and we would encounter naked
singularities. Excluding the latter we find that the kink is a
solution of lowest energy. Of course, the condition is automatically
satisfied for all horizons of the kink if it is satisfied for one of them.

Obviously the above requirement is easily
translated into the condition
\begin{equation}
  \label{eq:2.6}
  W(\phi) = -2 w(\phi)^2\ ,
\end{equation}
where the numerical factor has chosen in such a way that it coincides
with \eqref{eq:2.1}. Not surprisingly we find that all BPS kinks are
stable (for the specific example of the kink of ref.\ \cite{Guralnik:2003we}
this has been observed in ref.\ \cite{Bergamin:2004me} already). But the
converse is true as well: Any stable kink solution allows a
supersymmetric extension and therein it is BPS. Accordingly all
horizons of a stable kink are extremal and therefore the
thermodynamical stability considered above is automatically guaranteed.

\section{Polynomial Potentials}
\label{sec:polynom}

To illustrate the general features of the kinks derived in the
previous section we present a few examples. As a first simplification
we set the conformal factor to unity ($e^Q =1$, $Z=0$), which does not
change any important physical behavior of the kink\footnote{It is
  important to notice that conformal transformations do change the
  physics, if they are not globally well defined. However, such
  conformal transformations would destroy the kink
  \cite{Bergamin:2004me}.}. Furthermore we simplify our considerations
by choosing polynomials with a maximal number of real roots. By rescaling
 the dilaton field, the coefficient of the leading
term can be set to $\pm 1$ and furthermore the shift-invariance is
used to set $\phi_F=-\phi_I=\phi_A$. We can make an ansatz of this
type either for $V$ or for $W$. Potentials $V$ of odd degree describe dS-dS (``+''-sign) or AdS-AdS (``-''-sign)
kinks, even degrees dS-AdS or AdS-dS kinks resp. We restrict our
discussion to the case with the negative sign, as this situation alone
can yield stable kink solutions and thus be extended to supergravity. All results on the bosonic kinks
generalize straightforwardly to the positive sign.
The choice 
\begin{equation}
  \label{eq:3.1}
  V = - (\phi+\phi_A)(\phi-\phi_2)\ldots(\phi-\phi_{n-1})(\phi-\phi_A)
\end{equation}
has the advantage that all CDVs together with the associated curvatures
\begin{equation}
  \label{eq:3.1.1}
  R(\phi_i) = - 2 \prod_{j\neq i} (\phi_i-\phi_j)
\end{equation}
are immediate. 
However the positions of the horizons in
general cannot be determined analytically. This may be seen as a minor
problem for the interior horizons, but if the condition
$W(\phi_A)=W(-\phi_A)$ cannot be solved we are unable to determine
whether the potential \eqref{eq:3.1} actually has a kink solution or not. There
exists one class of potentials that does not suffer of this problem,
the symmetric AdS-AdS (or dS-dS) kink. It is described by
    \begin{equation}
      \label{eq:3.1.2}
      V = -
      (\phi+\phi_A)(\phi+\phi_2)\ldots(\phi+\phi_n)\phi(\phi-\phi_n)\ldots(\phi-\phi_A)\ ,
    \end{equation}
which automatically satisfies $W(\phi_A)=W(-\phi_A)$. The relation
\eqref{eq:3.1.1} simplifies to
\begin{align}
  \label{eq:3.1.3}
      R(\phi_i) &= R(-\phi_i)= - 4 \phi_i^2 \prod_{j\neq i} (\phi_i^2 -\phi_j^2)\ , &
      R(0) &= 2 (-1)^{n+1} \prod_i \phi_i^2\ .
\end{align}
Unfortunately we are not able to determine the number and the positions
of interior horizons. As is seen from eq.\ \eqref{eq:1.9} even a large
number of CDVs cannot guarantee the existence of interior horizons.

Instead of \eqref{eq:3.1.2} the more restrictive ansatz
\begin{equation}
  \label{eq:3.1.4}
  W = -(\phi+\phi_A)(\phi-\phi_2)\ldots(\phi-\phi_{n-1})(\phi-\phi_A)
\end{equation}
can be made. Here $\phi_A$ still refers to the ``asymptotic'' CDVs as
in \eqref{eq:3.1}, but the other $\phi_i$ now label the positions of
the horizons instead of the CDVs. Therefore this ansatz yields $2n$
horizons while the corresponding $V$ generates $2n-1$ CDVs, which
corresponds to the equality in eq.\ \eqref{eq:1.9}. This way the causal
structure of the solution becomes obvious however the positions and
characteristics (curvatures) of the different CDVs are not available
in general.

Also in supersymmetric theories two different ans\"atze can be made, either for the prepotential
$w$ or its derivative. By choosing
\begin{equation}
  \label{eq:3.1.5}
  w = (\phi+\phi_A)(\phi-\phi_2)\ldots(\phi-\phi_{n-1})(\phi-\phi_A)
\end{equation}
with eq.\ \eqref{eq:2.1} the supersymmetric version of eq.\ \eqref{eq:3.1.4} is
obtained. According to the discussion above the
$\phi_i$ label the horizons and at the same time all $w=0$
CDVs. They are characterized by the curvatures
\begin{equation}
  \label{eq:3.1.6}
  R(\phi_i) = - 4 \prod_{j\neq i} (\phi_i-\phi_j)^2\ .
\end{equation}
All kink solutions from eq.\ \eqref{eq:3.1.5} are stable and within
supergravity they are BPS states.
A non-BPS kink is possible only if $w$ does not have the maximal
number of real roots, which is always possible according to
eq.\ \eqref{eq:2.3}.

\subsection{The Simplest Kink}
\label{sec:simplest}
For illustrational purposes potentials of order three, four and five
are considered more in detail.
A kink solution requires at least three CDVs, therefore the simplest
kink is described by the potential
\begin{equation}
  \label{eq:3.2}
  V = - (\phi+\phi_A)(\phi-\phi_2)(\phi-\phi_A)\ .
\end{equation}
The existence of a kink solution requires $W(\phi_A)=W(-\phi_A)$ and
therefore $\phi_2=0$. By an appropriate choice of the integration
constant in eq.\ \eqref{eq:1.6} $W$ can be written as
\begin{equation}
  \label{eq:3.3}
  W = -\inv{4} \phi^4 + \half{1} \phi^2_A \phi^2 - \inv{4} \phi^4_A =
  -\inv{4} (\phi^2 - \phi_A^2)^2\ .
\end{equation}
This is exactly the kink found in refs.\ \cite{Guralnik:2003we,Grumiller:2003ad}, to this
order the kink is unique. All models allow a supersymmetric extension
and the kink is BPS therein \cite{Bergamin:2004me}. For any further
discussions we refer to refs.\
\cite{Guralnik:2003we,Grumiller:2003ad,Bergamin:2004me}. 

\subsection{The Simplest AdS-dS kink}
\label{sec:adsds}

An AdS-dS kink requires at least four CDVs and has at least one horizon
in its interior. The simplest potential is thus
\begin{equation}
  \label{eq:3.4}
  V = - (\phi+\phi_A)(\phi-\phi_2)(\phi-\phi_3)(\phi-\phi_A)\ .
\end{equation}
A kink solution exists if $\phi_2 \phi_3 = - \phi^2_A/5$ and therefore
$\phi_2<0$, $\phi_3>0$. A more stringent restriction comes from the
fact that $\phi_2$ and $\phi_3$ shall lie between $-\phi_A$ and
$\phi_A$, thus
\begin{align}
\label{eq:3.5}
\phi_2 &= - \inv{5} \frac{\phi_A^2}{\phi_3}\ , & \phi_A \geq \phi_3 \geq
\inv{5} \phi_A\ .
\end{align}
Again we choose the integration constant in eq.\ \eqref{eq:1.6} in such a
way that the kink has $C=0$. Then it is easily seen that $W(\phi)$
must have five zeros, namely two at $\phi=-\phi_A$ and at
$\phi=\phi_A$ resp., and one at the interior horizon $\phi=\phi_h$. Accordingly $W$ can be written as
\begin{align}
  \label{eq:3.5.1}
  W &= - \inv{5} (\phi-\phi_A)^2 (\phi+\phi_A)^2 (\phi-\phi_h)\ , &
  \phi_h &= \inv{4 \phi_3} (5 \phi_3^2 - \phi_A^2)\ .
\end{align}
There exist three special choices of $\phi_3$:
\begin{description}
\item[\boldmath$\phi_3 = \phi_A/\sqrt{5}$:] This is the
  ``symmetric'' kink, $-\phi_2 = \phi_3 = \phi_I$, $|R_A|/|R_I| =
  \sqrt{5}$, the interior horizon is at $\phi_h=0$.
\item[\boldmath$\phi_3 = \phi_A$, $\phi_3 = \phi_A/5$:] One of the interior
  CDVs and the interior horizon coincide with $\phi_A$ or $-\phi_A$ resp., $W$ has a triple zero there. Thus the
  AdS CDV ($\phi_3 = \phi_A$) or the dS CDV ($\phi_3 = \phi_A/5$)
  turn into Minkowski/Rindler space ($R=0$). 
\end{description}

\subsection{An Extended Kink}
\label{sec:extended}
The first two examples provided the simplest possibilities for an
AdS-AdS (dS-dS) and a AdS-dS kink, resp. Our last example concerns a
non-minimal realization of the AdS-AdS kink with the ``minimal
non-minimal'' ansatz
\begin{equation}
  \label{eq:3.6}
  V = -
  (\phi+\phi_A)(\phi-\phi_2)(\phi-\phi_3)(\phi-\phi_4)(\phi-\phi_A)\ .
\end{equation}
The condition $W(\phi_A) = W(-\phi_A)$ yields the constraint
\begin{equation}
  \label{eq:3.7}
  \inv{5} (\phi_2 + \phi_3 + \phi_4) \phi_A^2 = - \phi_2\phi_3\phi_4\ .
\end{equation}
There exist three possibilities for the causal structure of this
solution: The kink can have no interior horizon
($W(\phi_3)<W(\phi_A)$), two interior horizons ($W(\phi_3)>W(\phi_A)$)
or one extremal interior horizon ($W(\phi_3)=W(\phi_A)$). For solutions
with interior horizons $W$ has the maximal number of six real roots and the
positions of the horizons $\phi_{h_1}$ and $\phi_{h_2}$ are again
straightforwardly deduced as
\begin{align}
  \label{eq:3.8}
  \phi_{h_1}+\phi_{h_2} &= \frac{6}{5}(\phi_2 + \phi_3 + \phi_4)\ , &
  \phi_{h_1}\phi_{h_2} &= \half{1} \phi_A^2 + \half{3} (\phi_2 \phi_3 +
  \phi_2 \phi_4 + \phi_3 \phi_4)\ .
\end{align}
Restricting to stable or equivalently supersymmetric solutions we
notice that $w$ is a polynomial of order three. Thus it can have
either one or three real roots, the former case obviously does not
lead to kink solutions. Therefore we encounter one extremal horizon in
the interior of the kink, which is obtained from eq.\ \eqref{eq:3.8}
by imposing  $\phi_{h_1}=\phi_{h_2}$, which consequently is also the position of
  $\phi_3$. The conformally invariant potential becomes $W = \bigl((\phi+\phi_A) (\phi-\phi_3) (\phi-\phi_A) \bigr)^2$,
  while $V$ is given by \eqref{eq:3.6} with
  \begin{align}
    \label{eq:3.12}
    \phi_2 \phi_4 &= - \frac{\phi^2_A}{3}\ , & \phi_3 &= \frac{3}{2} (\phi_2 +
    \phi_4)\ .
  \end{align}
  Of course, the special choice $\phi_4 = \phi_A$ implies $\phi_3 =
  \phi_A$ and $W$ has a fourth order root at $\phi_A$. Thus this
  ansatz also provides the simples example for a stable kink with one
  $R=0$ CDV. Finally we note the simplest stable $(R=0)$-$(R=0)$ kink
  needs a $w$ of order four, which is at the same time the simplest
  example where the inequality \eqref{eq:2.3} needs not be
  satisfied. It is seen that interesting examples of non-minimal kinks
  immediately lead to polynomials of order higher than four, such that
  a general analytic treatment is no longer possible.

\section{Conclusions}
\label{sec:conclusions}
We have presented the systematics of constant dilaton vacua and the
associated kink solutions, originally motivated from the KK reduced
gravitational Chern-Simons term
\cite{Guralnik:2003we,Grumiller:2003ad}, in 2D generalized dilaton
gravity and supergravity. In order that such solutions exist various
conditions must be satisfied. This concerns the
existence of constant dilaton vacua (they occur at the roots of the
potential $V$, only), but even more importantly a kink can exist only
if the value of the conformally invariant potential is the same for
the initial and final point of the kink. Restrictions that emerge from
supergravity have been discussed and it has been found that all stable
kink solutions allow a supersymmetric extension and are BPS therein. 

At hand of the $N=(2,2)$ supersymmetric case it has been shown that
an extension of the (scalar) field content does not necessarily lead
to a richer structure of kinks. Nevertheless, models with
complex dilaton could be an especially interesting topic to consider, if space-time is
complexified as well \footnote{Complexified space-time in 2D dilaton gravity
was considered in \cite{Bergamin:2004pn}, however the dilaton remained a real field in
that approach.}.
There exist other interesting extensions of this work. Besides
deeper investigations of specific examples many questions of
thermodynamics remain open. This concerns the faith of unstable kink
configurations as well as problems raised in ref.\ \cite{Bergamin:2004me}
already, that can be discussed in a broader context now.

Finally the question of asymptotics should be mentioned, discussed
in detail in ref.\ \cite{Grumiller:2003ad} for the specific example of the
kink in ref.\ \cite{Guralnik:2003we}. As is obvious from its
definition given here, the initial and final points of the kink cannot
represent spacial infinity. To overcome this problem a patching of
(A)dS vacua has been proposed in ref.\ \cite{Grumiller:2003ad}, which
however induces matter fluxes at the initial and final horizons. Another
way out has been presented in ref.\ \cite{Bergamin:2004me}. In any way it
should be kept in mind that the kink itself does not represent a
global solution of 2D dilaton gravity and thus must be completed in an
appropriate way, which of course may depend on the specific application.

{\bf Acknowledgment.}  It is a pleasure to acknowledge stimulating
discussions with D.~Grumiller, including a collaboration at an early
stage of this work. This work has been supported by the project
P-16030-N08 of the Austrian Science Foundation (FWF). I would like to
thank the organizers of GAS@BS 05 for all their work that made this
interesting and fruitful workshop possible.
I am especially indebted to R.\ and M.\ Poncet for a wonderful stay
in Termes (Ard\`eche, FR), where most of this work has been developed.

\end{document}